\documentclass{vgtc}                          

\usepackage{balance}       
\usepackage{graphics}      
\usepackage{color}
\usepackage{booktabs}
\usepackage{textcomp}
\usepackage{enumerate}
\usepackage[table]{xcolor}
\usepackage{lipsum}
\usepackage{makecell}
\usepackage{multicol}
\usepackage{multirow}
\usepackage{array}
\usepackage{verbatimbox}
\usepackage{enumitem}
\usepackage{amsmath}
\usepackage{stfloats}
\usepackage{amsthm}
\usepackage{listings}
\usepackage{caption} 
\usepackage{xspace}
\usepackage[linesnumbered]{algorithm2e}
\usepackage{algpseudocode}
\usepackage{tabularx}
\usepackage{arydshln}
\usepackage[bottom]{footmisc}
\usepackage{stackengine}
\usepackage{placeins}
\usepackage{graphicx}
\usepackage{subcaption} 
\usepackage{pifont}
\usepackage{tcolorbox}
\usepackage{colortbl}
\usepackage{wrapfig}

\newcommand*{\eg}{\textit{e.g.},\xspace}
\newcommand*{\ie}{\textit{i.e.},\xspace}
\newcommand*{\vs}{\textit{vs.}\xspace}

\newcolumntype{L}[1]{>{\raggedright\let\newline\\\arraybackslash\hspace{0pt}}m{#1}}
\newcolumntype{C}[1]{>{\centering\let\newline\\\arraybackslash\hspace{0pt}}m{#1}}
\newcolumntype{R}[1]{>{\raggedleft\let\newline\\\arraybackslash\hspace{0pt}}m{#1}}

\makeatletter
\def\thickhline{%
  \noalign{\ifnum0=`}\fi\hrule \@height \thickarrayrulewidth \futurelet
   \reserved@a\@xthickhline}
\def\@xthickhline{\ifx\reserved@a\thickhline
               \vskip\doublerulesep
               \vskip-\thickarrayrulewidth
             \fi
      \ifnum0=`{\fi}}
\makeatother

\makeatletter
\def\thickhlinespace{%
  \addlinespace[1ex]
  \noalign{\ifnum0=`}\fi\hrule \@height \thickarrayrulewidth \futurelet
   \reserved@a\@xthickhline
   \addlinespace[1ex]
   }
\def\@xthickhlinespace{\ifx\reserved@a\thickhline
               \vskip\doublerulesep
               \vskip-\thickarrayrulewidth
             \fi
      \ifnum0=`{\fi}}
\makeatother

\newlength{\thickarrayrulewidth}
\newlength\Origarrayrulewidth

\algnewcommand{\IfThenElse}[3]{
  \State \algorithmicif\ #1\ \algorithmicthen\ #2\ \algorithmicelse\ #3}

\newcommand{\statMean}[1]{\textit{Mean} = {#1}}
\newcommand{\statSD}[1]{\textit{SD} = {#1}}
\newcommand{\statBeta}[1]{$\beta$ = {#1}}
\newcommand{\statT}[2]{\textit{t}({#1}) = {#2}}
\newcommand{\statZ}[1]{\textit{z} = {#1}}
\NewDocumentCommand{\statP}{O{=} m}{%
  \textit{p} #1 {#2}%
}
\newcommand{\statCI}[2]{\textit{CI}[{#1}, {#2}]}

\definecolor{darkgrey}{HTML}{7C7C7C}
\newcommand{\mean}[1]{\textcolor[HTML]{74A439}{\textit{#1}}}
\newcommand{\CI}[1]{\textcolor[HTML]{DDAD3B}{\textit{#1}}}
\newcommand{\MFV}[1]{\textcolor[HTML]{4B9C7A}{\textit{#1}}}
\newcommand{\violin}[1]{\textcolor[HTML]{9F7831}{\textit{#1}}}
\newcommand{\density}[1]{\textcolor[HTML]{7470AE}{\textit{#1}}}
\newcommand{\HMFV}[1]{\textcolor[HTML]{CA6628}{\textit{#1}}}
\newcommand{\PMFV}[1]{\textcolor[HTML]{D53E88}{\textit{#1}}}
\newcommand{\BMFV}[1]{\textcolor[HTML]{666666}{\textit{#1}}}

\newcommand{\bfsection}[1]{\noindent \textbf{{#1}}~}

\definecolor{revisionColor}{HTML}{025DF4}

\usepackage{mdframed}
\newcommand{\infobox}[2]{%
\begin{center}
\begin{minipage}{\linewidth}
\begin{mdframed}[
    linewidth=0.8pt,
    backgroundcolor=gray!10,
    linecolor=black,
    innerrightmargin=6pt,
    innerleftmargin=6pt,
    innertopmargin=4pt,
    innerbottommargin=4pt
]
\textbf{#1:} #2
\end{mdframed}
\end{minipage}
\end{center}
}

\graphicspath{{figures/}{pictures/}{images/}{./}} 

\usepackage{times}                     

\usepackage{tabu}                      
\usepackage{booktabs}                  
\usepackage{lipsum}                    
\usepackage{mwe}                       

\usepackage{mathptmx}                  

\onlineid{9372}

\vgtccategory{Research}

\vgtcinsertpkg

\title{Striking a Balance: Evaluating How Aggregations of Multiple Forecasts Impact Judgment Under Uncertainty}

\author{
Ruishi Zou\thanks{This work was done while he was a visiting student at the Northeastern University. E-mail: ruzou@ucsd.edu}\\ %
        \scriptsize University of California, San Diego %
\and Siyi Wu\thanks{e-mail: reyna.wu@mail.utoronto.ca}\\ %
     \scriptsize University of Toronto %
\and Racquel Fygenson\thanks{e-mail: fygenson.r@northeastern.edu}\\ %
     \scriptsize Northeastern University
\and Bingsheng Yao\thanks{e-mail: b.yao@northeastern.edu}\\ %
     \scriptsize Northeastern University
\and Dakuo Wang\thanks{E-mail: d.wang@northeastern.edu}\\ %
     \scriptsize Northeastern University
\and Lace Padilla\thanks{E-mail: l.padilla@northeastern.edu}\\ %
     \scriptsize Northeastern University
}

\teaser{
  \centering
  \vspace{-15px}
  \includegraphics[width=0.86\linewidth]{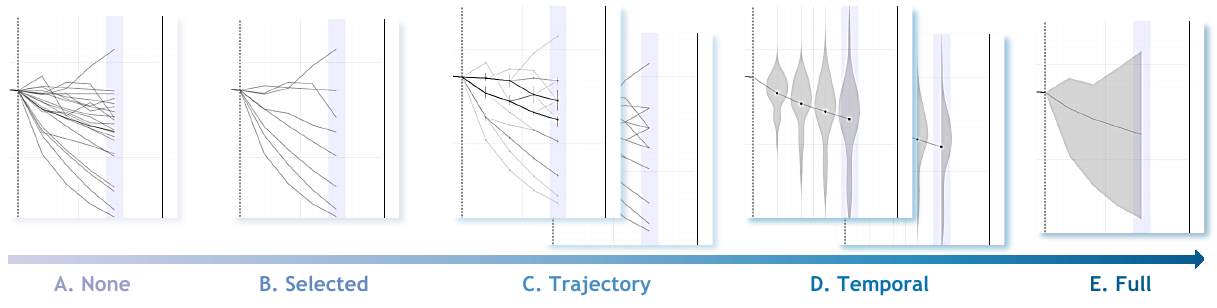}
  \caption{Five \textbf{aggregation levels} of visualizations for multiple forecasts. The aggregation level \textit{increases} from \textbf{A} to \textbf{E}: \textbf{A} represents all individual forecasts, while \textbf{E} represents a statistical summary of all forecasts.}
  \label{fig:teaser}
}

\abstract{
    
Decision-makers consult multiple forecasts to account for uncertainties when forming judgments about future events. While prior works have compared unaggregated and highly-aggregated designs for displaying multiple forecasts (\eg Multiple Forecast Visualizations versus confidence interval plots), it remains unclear how \textit{partial} aggregation impacts judgment. To investigate the effect of partial aggregation, we curated three designs that partially aggregate multiple forecasts. 
Through two large-scale studies (Experiment 1 \textit{n} = 695 and Experiment 2 \textit{n} = 389) across 14 judgment-related metrics, we observed that one design (Horizon Sampled MFV) significantly enhanced participants' ability to predict future trends, thereby reducing their surprise when confronted with the actual outcomes. Grounded in empirical evidence, we provide insights into how to design visualizations for multiple forecasts to communicate uncertainty more effectively. Specifically, since no approach excels in \textit{all} metrics, we advise choosing different designs based on communication goals and prior knowledge of forecasts.

} 


\begin{document}


\maketitle

\section{Introduction}
\label{sec:introduction}

Making judgments about future events is notoriously challenging, partly because humans struggle with understanding uncertainty~\cite{tversky1974}. As computational forecast models become more prevalent, one promising approach for communicating uncertainty of future events is through multiple forecasts. The multiple forecasts approach involves aggregating predictions from independent groups to provide chart readers (\eg the general public, experts) with a comprehensive overview of potential future outcomes. Specifically, examples of multiple forecasts can be seen in COVID-19 projections~\cite{reich2023visualization}, climate modeling~\cite{2024cfsv2}, and %
weather forecasts~\cite{2025breezyweather, 2025free}.

Despite this growing prevalence, communicating uncertainty across multiple forecasts remains challenging. According to Sarma et al.~\cite{sarma2025more}, multiple forecasts contain two types of uncertainty: \textit{probabilistic uncertainty} (quantified uncertainty that can be represented probabilistically) and \textit{incertitude} (uncertainty about how well each model represents reality). Because readers confront uncertainty types when forming judgments about the future~\cite{sarma2025more}, visualizations of multiple forecasts should represent both uncertainties.

Although prior work has investigated how different visualizations of multiple forecasts~\cite{padilla2022impact, padilla2023multiple} and multiple model outputs~\cite{sarma2025more} impact judgment and decision-making, these studies typically employ methods at extreme ends of the visual aggregation spectrum: showing all forecasts and \textbf{no aggregation} (\eg ensembles~\cite{sarma2025more}, Multiple Forecast Visualizations or MFVs~\cite{padilla2023multiple}), or showing a statistical summary of all forecasts with \textbf{full aggregation} (\eg \textit{probability boxes}~\cite{sarma2025more}, \textit{confidence intervals},~\cite{padilla2023multiple}). The effect of intermediate approaches, such as \textbf{partial aggregation} of multiple forecasts, remains unclear~\cite{sarma2025more}. Motivated by this gap, we ask: \textbf{\textit{How do different aggregation levels of multiple forecasts impact readers' judgment of future events under uncertainty?}}

To answer this question, we conducted two large-scale online studies evaluating five design categories (Fig.~\ref{fig:teaser}, A-E): \textit{none} (show all forecasts), \textit{selected} (aggregate representative groups), \textit{trajectory} (aggregate by trajectory), \textit{temporal} (aggregate by time points), and \textit{full} (show summary forecast). In Sec.~\ref{sec:alternativeMTFV}, we explain our approach for curating visualizations using partial aggregation. In Sec.~\ref{sec:experiment-design},  we describe how we investigated these designs through two evaluations (Experiment 1, \textit{n} = 695; Experiment 2, \textit{n} = 389). To ensure ecological validity, we used a real-world dataset, COVID-19 Forecast Hub~\cite{cramer2022uniteda}, to generate our study stimuli. We selected five sub-datasets that have 1) a varying forecast-outcome relationship (\eg aligned, contrasting), and 2) different numbers of forecasts to approximate a broader set of real-world datasets. Moreover, we evaluate reader performance across 14 metrics that reflect judgment processes, including judgment performance in predicting future outcomes~\cite{padilla2023multiple}, trust~\cite{yang2023swaying}, surprise in response to the actual outcome~\cite{yang2023subjective}, and cognitive effort~\cite{hart1988development, castro2022examining}. In this context, we define readers as the general public, because COVID-19 visualizations were primarily developed for public-facing communication~\cite{zhang2021mapping}. 

Through our analysis in Sec.~\ref{sec:results}, we found that no single approach outperforms others in all metrics. Showing a selected subset of multiple forecasts (Fig.~\ref{fig:teaser}.B) is the best approach for eliciting accurate future judgments and incurring minimal surprise. However, showing a more aggregated version (Fig.~\ref{fig:teaser}.D or E) might be the best approach to elicit trust. In Sec.~\ref{sec:discussion}, we outline design recommendations for visualizing multiple forecasts with different communication goals (\eg eliciting trust) and forecast incertitudes (\eg the known accuracy of forecasts). In summary, we contribute:%

\begin{itemize}
[leftmargin=12pt, itemsep=0pt, parsep=0pt, partopsep=0pt, topsep=0pt]
    \item Three partial aggregation methods visualizing multiple forecasts
    \item An investigation of how different aggregation levels 
    impacts readers' judgment viewing multiple forecasts
    \item Empirical evidence for a correlation between different aggregation levels of multiple forecasts and facets of judgment
    \item Advice for designers visualizing multiple forecasts on their communication goals and knowledge of the forecast models
\end{itemize}

\section{Related Work}
\label{sec:related_work}

We situate our work within three lines of prior research: 1) uncertainty visualization methods,  %
2) visualizing multiple forecasts, %
and 3) evaluation metrics for uncertainty visualization.%

\subsection{Uncertainty Visualization}
\label{subsec:uncertainty-design-space}

Most visualization designers choose between two approaches in presenting uncertainty: using intervals or ratios to present the summary statistics or using distributions to present a more expressive representation of uncertainty~\cite{padilla2021uncertainty}. Although using confidence intervals or box plots remains a popular approach to present uncertainty~\cite{zhang2021mapping}, past research indicates that chart readers can misinterpret the distributional information in the summaries as categorical~\cite{belia2005researchers, joslyn2013error, correll2014error, padilla2017effects, hofman2020how, joslyn2020, castro2022examining, kale2021visual}. 

A growing body of research advocates for distributional visualizations, such as violin~\cite{hintze1998violin, correll2014error}, density~\cite{green1985comparison}, gradient~\cite{correll2014error}, strip~\cite{kay2016when}, and ensemble plots~\cite{liu2017uncertainty, liu2019visualizing}, as more effective methods for displaying distributional data. Although prevailing studies underscore the superiority of distributional visualizations, the specific contributors to their enhanced effectiveness remain debatable. Meanwhile, there are also limitations, with studies revealing categorical biases in hurricane forecasting ensembles~\cite{padilla2017effects, padilla2020powerful}. 

Other research has suggested using frequency framing (\eg ``5 out of 10'' \vs ``50\%'') to present uncertainty in a way that aligns with everyday experiences~\cite{kay2016when, pu2020probabilistic, fernandes2018uncertainty, kale2019hypothetical}. One example of frequency framing in visualization is the Hypothetical Outcome Plot (HOP)~\cite{hullman2015hypothetical, kale2019hypothetical}, which displays a different possible outcome in each frame of an animation to induce its readers to form a mental model of a forecast's underlying distribution. Empirical studies suggest that HOPs improve the general public's distribution-based judgments~\cite{hullman2015hypothetical, kale2019hypothetical, kim2019bayesian}, highlighting the potential of using frequency framing to communicate uncertainty to the general public.

\subsection{Visualizing Multiple Forecasts}
\label{subsec:multiple-forecasts}

Multiple outputs from \textit{one forecast} have been commonly visualized through two ways: ensemble visualizations (for review see~\cite{wang2018visualization}) or summary representations (\eg confidence intervals, probability boxes~\cite{sarma2025more}, contour box plots~\cite{whitaker2013contour}). 
Recent work has shifted focus toward methods for representing multiple predictions from \textit{independent forecasts} (multiple forecasts), rather than outputs from a single model (multiple outputs). MFVs look similar to ensemble visualizations but reveal distributional information of multiple forecasts by showing the independent forecasts in a single plot~\cite{padilla2017effects, padilla2020powerful}.

While these works contribute to our understanding of the effects of visualizing multiple forecasts, numerous unresolved design questions surround visualizing multiple forecasts exist. In this paper, we investigate how the degree of aggregation of individual forecasts impacts chart readers' judgments. We motivate the levels of aggregation question given the ever-growing sizes of forecast collections (\eg, COVID-19 Forecast Hub~\cite{cramer2022uniteda} has over 50 models), and visualizing all models simultaneously would result in substantial overplotting. To conduct partial aggregation of forecasts, we designed three heuristically inspired approaches for partially aggregating multiple forecasts and evaluated the design against two typical forecast visualization approaches (confidence interval and MFV). Through our investigation, we aim to explore if 1) we can ``strike a balance'' in the degree of aggregation, and 2) how to best design the visualization of multiple forecasts given different communication goals and prior knowledge of the forecasts' models.

\subsection{Evaluation Measures for Uncertainty Visualization}
\label{subsec:eval-uncertainty-decision}
Researchers have applied various metrics to evaluate the utility of uncertainty visualizations (for a review, see~\cite{hullman2019pursuit}). Many studies evaluate \textit{performance}, or participants' ability to complete specific tasks using a given visualization. For example, researchers have used \textit{absolute error}, or the numerical deviation from a ground truth value, to present judgment performance~\cite{hullman2015hypothetical, zhao2023evaluating}. Others have used the concept of \textit{accuracy}, or the percentage of matches with a categorical truth~\cite{xiong2022reasoning}. Besides evaluating a single response, researchers have also explored using the \textit{selected range} to quantify readers' judgments~\cite{gschwandtnei2016visual}. Beyond performance-based measures, self-reported measures, including \textit{trust}~\cite{mayr2019trust, zhao2023evaluating, padilla2023multiple, yang2023subjective, yang2020how, kim2021bayesianassisted, kong2019trust}, \textit{surprise}~\cite{yang2023swaying}, and \textit{perceived effort}~\cite{castro2022examining, riveiro2016visually, ranasinghe2019visualising}, have been used to evaluate uncertainty representations. %
For instance, researchers have used trust games~\cite{gaba2023my}, and single-item~\cite{kim2021bayesianassisted, padilla2023multiple} and multi-item trust scales~\cite{yang2023subjective, elhamdadi2024vistrust} to evaluate self-reported measures.

In this study, we consider both performance and self-reported measures to provide a holistic evaluation of how different aggregation levels of multiple forecasts impact readers' judgments. Specifically, we focus on four facets that reflect the readers' ability to make informed judgments: performance, trust, surprise, and effort. We outline the specific derivation of the measures in Sec.~\ref{subsec:measures}.

\section{Design Space: Aggregating Multiple Forecasts}
\label{sec:alternativeMTFV}

Before investigating how different aggregation levels of multiple forecasts impact judgment under uncertainty, we developed a set of designs that partially aggregate multiple forecasts. In the following, we describe a heuristic aggregation method inspired by a sampling method (Sec.~\ref{subsec:aggregation-sampling}), as well as other aggregations motivated by uncertainty visualization literature (Sec.~\ref{subsec:design-space}).

\subsection{Aggregation through Sampling}
\label{subsec:aggregation-sampling}
Showing samples of a larger corpus of data is a widely applied visualization technique to create an aggregated view of underlying data~\cite{ramosrojas2017sampling, yuan2021evaluation, liu2017uncertainty}. However, some forecast sampling methods require model metadata or bespoke visual encoding, making them application-specific and applicable only in professional settings. To ensure minimal data requirements and ease of interpretation, we adhered to two constraints when developing our first partially aggregated MFV design: 1) the sampling method should be a postprocessing technique and thus not need prior information on the forecasts, and 2) the visual encoding should be relatively common (\ie based on a line chart).

Considering the constraints, we selected DBSCAN~\cite{ester1996densitybased}, a density-based clustering algorithm, as our sampling technique for partially aggregating multiple forecasts. We made this decision based on two factors: 1) the algorithmic approach of DBSCAN is analogous to the human cognition process known as the Gestalt principle of proximity~\cite{todorovic2008gestalt}, and 2) DBSCAN-based sampling does not require forecast metadata, making it purely a postprocessing technique. In the following, we describe three designs at different aggregation levels using DBSCAN.

\subsubsection{Selected Forecasts: Horizon Sampled MFVs}

One direct approach for partial aggregation is displaying a subset of ``representative'' forecasts. In this aggregation process, we aimed to 1) reduce the number of forecasts such that it is within a recommended range of six to nine forecasts~\cite{padilla2023multiple}, and 2) maintain the shape and distribution of forecasts \textit{at the forecast horizon} (\ie the furthest predicted point in time). We call it \HMFV{Horizon Sampled MFV}. 

\begin{figure}[t]
    \centering
    \includegraphics[width=0.76\linewidth]{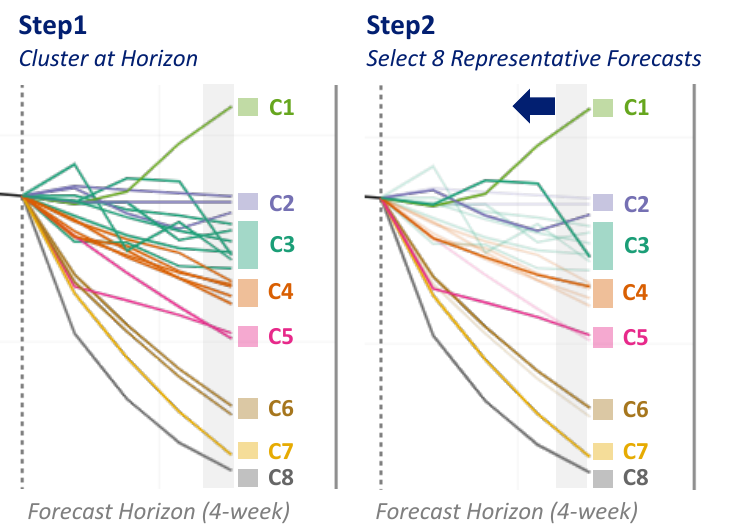}
    \caption{Generation process for a Horizon Sampled MFV. We encode clusters (C1-C8) generated at the forecast horizon with color and label the selected forecast path in dark blue.}
    \label{fig:smfv}
    \vspace{-8px}
\end{figure}

To generate a Horizon Sampled MFV, we first collected MFV forecasts into eight clusters (Fig.~\ref{fig:smfv}, left) using DBSCAN at the forecast horizon. We chose eight clusters as a middle point within the recommended range of six to nine forecasts. Next, we selected one forecast from each cluster to serve as a ``representative forecast'' (Fig.~\ref{fig:smfv}, right). %
Through the above two steps, we reduced the number of displayed forecasts to be within the recommended range, represented each forecast cluster equally, and maintained the shape and distribution of forecasts at the forecast horizon. Importantly, Horizon Sampled MFVs allow for single-forecast clusters, thus preserving forecasts that are outliers at the forecast horizon.

\subsubsection{Forecast Trajectories: Progressively Sampled MFVs}

To increase forecast aggregation, we extended our horizon sampling method to cluster at each forecast time point, not just the forecast horizon. This clustering method uncovers common trajectories and trends within multiple forecasts. This design was motivated by trend graphs~\cite{obermaier2016visual}, which apply a density-based clustering algorithm to group single-forecast ensemble data and then connect these clusters in a style analogous to Sankey diagrams~\cite{rosvall2010mapping} to depict time-varying ensembles. We adapted the trend graph design for multiple forecast visualizations by 1) clustering forecasts at time steps on the existing time-series coordinate plane, and 2) representing transitions between clusters with straight line segments. We name this partially aggregated MFV the \textit{Progressively Sampled MFV}.

\begin{figure}[t]
    \centering
    \includegraphics[width=0.88\linewidth]{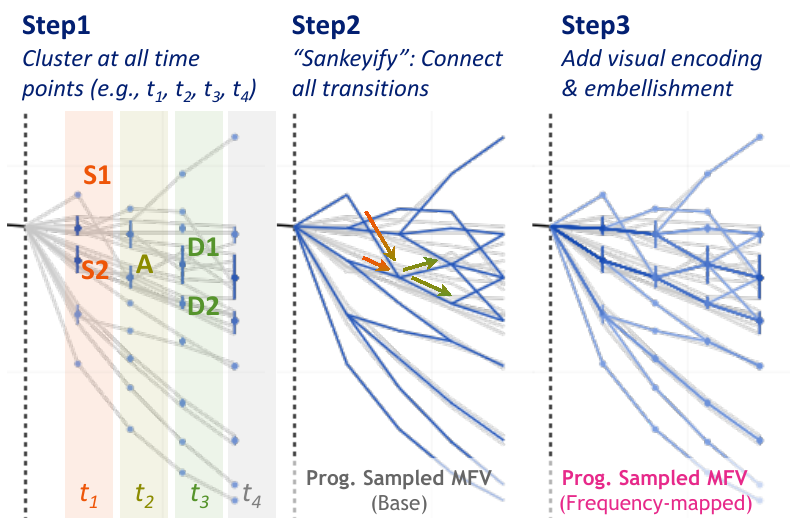}
    \caption{Generation process for Progressively Sampled MFVs. %
    In step 1, we cluster predictions at each time point ($t_1$-$t_4$). In step 2, we ``Sankify'' the clusters by connecting clusters where there exists an original forecast path. In step 3 we add visual encoding and embellishments to describe the number of forecasts represented by each cluster and the spread of each cluster's forecasts.}
  \vspace{-8px}
    \label{fig:htc}
\end{figure}

To generate a Progressively Sampled MFV, we first obtained clusters at each time point (Fig.~\ref{fig:htc},~step 1). We generated eight clusters at the forecast horizon (\ie $t_4$) to maintain consistency with the Horizon Sampled MFV. Then, we ``Sankeyified'' the original forecasts, connecting all possible transitions between clusters (Fig.~\ref{fig:htc},~step 2). For example, consider cluster \textit{A} at the second time point in Fig.~\ref{fig:htc}. We can observe that forecasts exist from source clusters \textit{S1} and \textit{S2}, directed towards destination clusters \textit{D1} and \textit{D2}, forming trajectories \textit{S1-A}, \textit{S2-A}, \textit{A-D1}, and \textit{A-D2}. We applied the same ``Sankeyify'' process to all the clusters from Step 1. We named these visualizations \BMFV{\textit{Base Progressively Sampled MFV}.}

To further improve the expressiveness of the Base Progressively Sampled MFV, we applied visual encoding and added visual embellishments to show cluster and trajectory information (Fig.~\ref{fig:htc},~step 3). Specifically, we used occlusion and opacity to depict each cluster's density (\ie number of forecasts) and the likelihood of trajectories (\ie number of forecasts in each transition). We also added a vertical bar to show the data range within the cluster (\ie more aggregated summary information). Since this design shows the frequency of forecasts existing in each cluster and trajectory, we name this design \PMFV{\textit{Frequency-mapped Progressively Sampled MFV}}.

In summary, Progressively Sampled MFVs are partial aggregations of MFVs, presenting multiple forecasts via individual clusters and their summary statistics.%

\begin{figure*}[t]
    \centering
    \includegraphics[width=.92\linewidth]{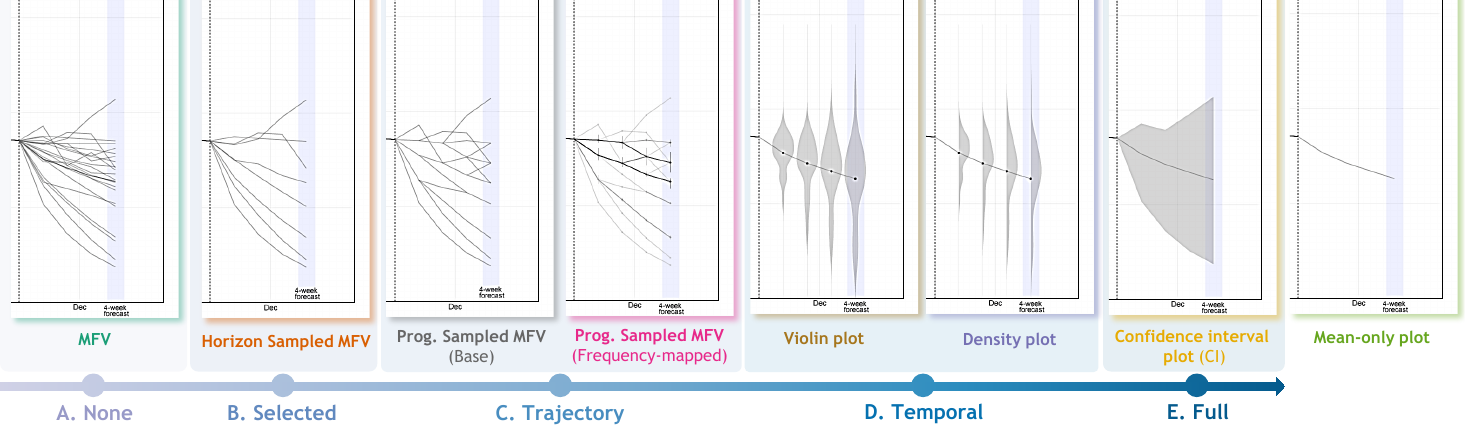}
    \caption{The visualizations mapped into the aggregation level spectrum: from \MFV{MFV} (no aggregation) to \CI{Confididence Interval Plot} (full aggregation). We also include a \mean{Mean-only Plot} to represent an option with no uncertainty representation.} %
    \label{fig:design-space}
      \vspace{-8px}
\end{figure*}

\subsection{Design Space: Spectrum of Aggregation Levels}
\label{subsec:design-space}

In addition to the three partially aggregated visualizations described above, we added five others to complete the aggregation design space (Fig.~\ref{fig:design-space}). We first selected the two extremes: \MFV{MFVs} show raw forecasts without aggregation, and \CI{Confidence Interval Plots} (CI plots) provide full aggregation across all forecasts and time points. We did not include a  gradient CI Plot (\ie gradient plot), which uses a color band to represent uncertainty~\cite{correll2014error} because non-gradient CI plots are more commonly used to visualize COVID-19 forecasts~\cite{zhang2021mapping}, and gradients can produce visual ``banding''.

Next, we outlined other forecast visualizations by their level of aggregation, between \MFV{MFVs} and \CI{Confidence Interval Plots}. \HMFV{Horizon Sampled MFVs} provide selective aggregation; \BMFV{Base Progressively Sampled MFVs} and \PMFV{Frequency-based Progressively Sampled MFV} offer trajectory-based aggregation with increasing summary information; \violin{Violin Plots} and \density{Density Plots} show statistical aggregation of all forecasts but do not aggregate across time points, and, lastly, \CI{Confidence Interval Plots} provide complete aggregation across all forecasts and time points. Additionally, we include \mean{Mean-only Plots} as control visualizations that depict no uncertainty.

It is worth noting that while \violin{Violin} and \density{Density Plots} are not conventional for displaying time-series data in line charts, we include them in this design space because they show distributional information of forecasts at each time point. This aggregation granularity is necessary between the trajectory level and the full level to bridge the gap between aggregating each time point as a whole or in clusters. For thoroughness, we included both the \violin{Violin} and \density{Density Plots}. Based on similar logic for avoiding gradient \CI{Confidence Interval Plots}, we opted to use shape (\ie violin, bell) to encode distributional information instead of color gradients.

\section{Methods and Experiment Design}
\label{sec:experiment-design}
To explore how different levels of aggregation impact multiple forecast readers' judgment under uncertainty, we conducted a comparative experiment that tested eight designs over 14 judgment-related metrics. In this section, we introduce our dataset (Sec.~\ref{subsec:dataset}), stimuli (Sec.~\ref{subsec:stimuli-gen}), experimental design (Sec.~\ref{subsec:experiment-design}), measures (Sec.~\ref{subsec:measures}), procedure and tasks (Sec.~\ref{subsec:procedure-tasks}), and participants (Sec.~\ref{subsec:participants}).

\subsection{Dataset}
\label{subsec:dataset}
We use the United States COVID-19 Forecast Hub dataset~\cite{cramer2022uniteda} to produce our experimental stimuli. We chose a real-world dataset rather than synthetic data to enhance the ecological validity of our study. %
The Forecast Hub is a central repository that stores COVID-19 indicators (\eg cases, hospitalization, mortality) and forecasts from over 50 agencies at a four-week forecast horizon. 

We chose mortality data to ensure the metrics visualized in our stimuli were intuitive %
for all participants. To avoid participants' past experiences with COVID-19 affecting their judgments, we omitted the term ``COVID-19'' and described the data as depicting mortality for ``a type of disease''. Additionally, we selected six sub-datasets %
(five for the study and one for the tutorial, see Tab.~\ref{tab:exp-dataset-choice}) based on the following criteria: 1) each sub-dataset was from a time point that contains multiple forecasts from distinct agencies, 2) the sub-dataset time points are sufficiently spaced to prevent any sub-datasets' forecasts from revealing another's actual outcomes, and 3) the sub-datasets represent a spectrum of truth-forecast relationships (\ie outlier, contrasting, and aligned scenarios).

We obtained the forecasts for the sub-datasets through the Forecast Hub dataset's Zoltar API at a four-week horizon. We filtered out models with the prefix \verb|COVIDhub| (\ie ensemble models) and those that had missing data. We use COVID-19 mortality data released by Johns Hopkins University as ground truth.

\begin{table}[!tb]
\small
\caption{List of the six sub-datasets we used to create our stimuli.}
\centering
    \begin{tabular}{cllll}
      \toprule
      \textbf{ID} & \textbf{Purpose} & \textbf{Date of Forecasts} & \textbf{\# Forecasts} & \textbf{Type} \\
      \midrule
      T1 & Tutorial & 2022-10-01 & 19 & -\\
      \midrule
      T2 & Study & 2020-11-14 & 43 & Outlier \\
      T3 & Study & 2021-01-02 & 39 & Contrast\\
      T4 & Study & 2021-05-08 & 44 & Aligned \\
      T5 & Study & 2021-11-13 & 24 & Contrast\\
      T6 & Study & 2022-04-16 & 30 & Contrast\\
      \bottomrule
    \end{tabular}
    \label{tab:exp-dataset-choice}
      \vspace{-8px}
\end{table}

\subsection{Stimuli}
\label{subsec:stimuli-gen}

We created 48 visualizations as stimuli combining the six forecast sub-datasets (Tab.~\ref{tab:exp-dataset-choice}) and eight designs (Fig.~\ref{fig:design-space}). To elicit responses to the self-reported measure of surprise, we subsequently crafted 40 visualizations from the five non-training sub-datasets that contained the actual outcome labeled as a red dot at a four-week horizon. %

\begin{figure}[t]
    \centering
    \includegraphics[width=0.8\linewidth]{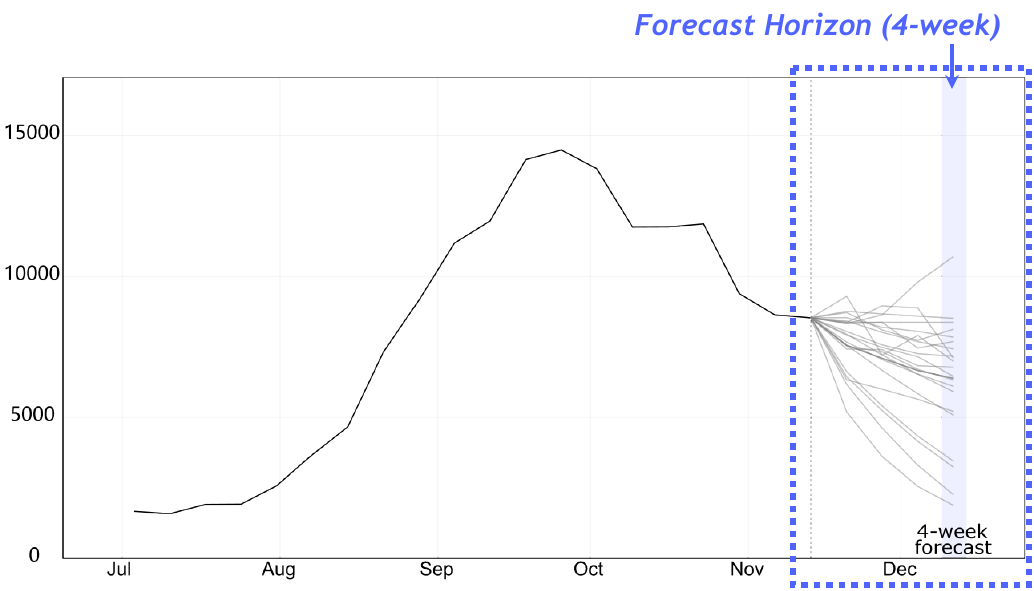}
    \caption{An example stimulus used in the study. The content in the purple box corresponds to a different visualization design.}
    \label{fig:stimuli-sample}
  \vspace{-8px}
\end{figure}

We generated all visualizations using the same visual style to minimize confounding variables. We used JavaScript packages, including \texttt{d3.js} v7.8.5 for visualization, \texttt{simple-statistics} v7.8.3 for statistical calculation, and \texttt{density-clustering} v1.3.0 for the DBSCAN algorithm. We rendered all visualizations with a base 2:1 aspect ratio, resulting in a final dimensions of $1290 \times 600$ px after we added axis annotations. To improve the clarity of the tasks, we added a light-blue band to indicate the forecast horizon (see Fig.~\ref{fig:stimuli-sample}). Then, we exported the images as JPEGs with Adobe Illustrator at 1500~\texttimes~767 px (96 PPI).

\subsection{Experimental Design}
\label{subsec:experiment-design}
\begin{figure*}[tb]
    \centering
    \includegraphics[width=\linewidth]{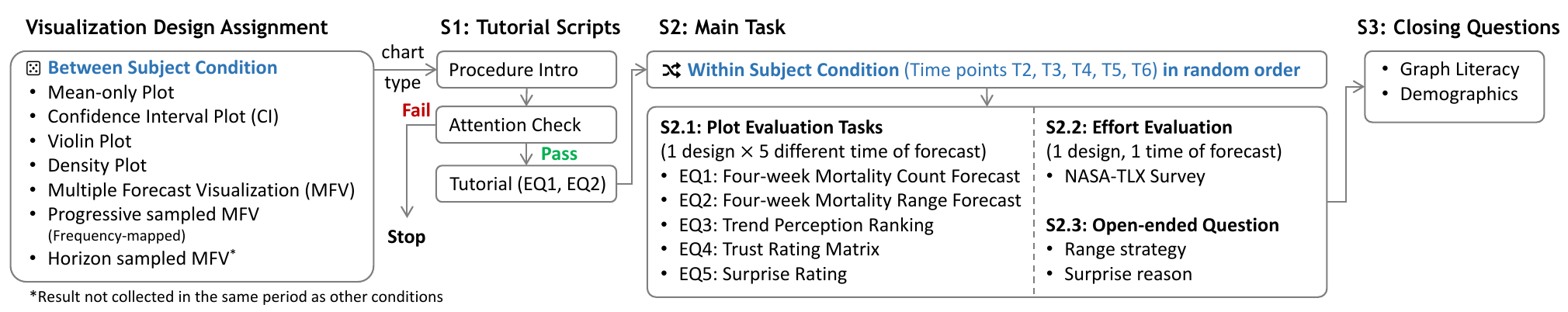}
    \caption{Experimental procedure of Experiment 1. Participants completed the tasks for one visualization design (between-subject condition) while responding to visualizations created from five sub-datasets %
    (within-subject condition, repeated measures) and answered five survey questions. The questions were presented in the order of \textbf{EQ1} to \textbf{EQ5}, with \textbf{EQ5} shown last to prevent disclosing the outcomes to participants.}
    \label{fig:experiment-procedure}
      \vspace{-8px}
\end{figure*}

Our study consists of two experiments. In \texttt{Experiment 1}, we tested the seven designs documented in Figure~\ref{fig:experiment-procedure} using a mixed between- and within-subjects design. %
Visualization type was the between-subjects factor, dividing participants into seven distinct groups. The within-subjects factor was sub-datasets (T2-T6 in Tab.~\ref{tab:exp-dataset-choice}), which we treated as a repeated measure to account for different data shapes. Participants completed five plot evaluation tasks (Fig.~\ref{fig:experiment-procedure}, S2) across the five sub-datasets, %
using the visualization design assigned to their group. This experimental setup resulted in a total of 25 trials per participant. The order of the forecast sub-datasets was randomized; however, within each sub-dataset, the sequence of questions remained consistent for all participants.

Initially, we created this experiment to test six visualization designs. However, after reviewing the results of this initial setup, we identified a gap between the no-aggregation MFV stimuli and the stimuli with aggregation that show MFVs' trajectories, and decided to add a seventh visualization design (Horizon Sampled MFV). This follow-up between-subjects condition was conducted online seven weeks after the initial sessions. An analysis of participant demographics suggested a potential shift in the population, introducing a confounding factor. To address this, we conducted a follow-up study (Experiment 2) with a single population to control for demographic variability and further examined the impact of different visualizations on readers' perceptions.

\texttt{Experiment 2} used the same study design as Experiment 1 but introduced a different set of between-subject conditions for validation and exploration. The visualization designs chosen for Experiment 2 included MFVs (referent), Horizon Sampled MFVs (for results validation), Base Progressively Sampled MFVs (for further exploring the trajectory aggregation level), and Violin Plots (for the temporal aggregation level). We further explain the rationale for choosing these four visualization designs in Sec.~\ref{sec:res-exp2}. Similar to Experiment 1, we treated Experiment 2's visualization designs as between-subject factors and the sub-datasets as repeated measures.

\subsection{Measures}
\label{subsec:measures}

We use four overall measures to evaluate users' judgment of multiple forecasts: performance, trust, surprise, and effort.

\paragraph{Performance}
We considered two direct measurements: \textbf{absolute error} and \textbf{range accuracy}. Because we rendered all visualizations at the same resolution, pixel values are directly comparable among stimuli. To measure judgment performance, we asked participants ``\textit{Based on this forecast, what do you think is the mortality count of this disease in the US in four weeks?}" and then instructed them to click on the visualization to record their answers. We then calculated the vertical pixel distance between the participants' response and the actual outcome to record \textbf{absolute error}.

Additionally, to evaluate how participants perceive the forecasts' spread, we prompted participants ``\textit{Based on this forecast, what do you think is the range of possible outcomes of this disease in the US in four weeks?}'' Participants were then instructed to click on the part of the visualization that corresponded to the upper and lower bounds of their estimates. We coded \textbf{range accuracy with a 0-1 binary code}, where 1 indicates that the actual outcome falls within the participant-specified upper and lower bounds, and 0 indicates otherwise. To avoid scenarios in which a wider participant response range may lead to more accurate range estimates, we introduced a third parameter, \textbf{range estimate} (distance between high and low bound) when we interpreted the results related to \textbf{range accuracy}.

\paragraph{Surprise}
We captured the degree of misalignment between the participant's mental model of the future outcome and the actual outcome, analogous to the concept of Bayesian surprise~\cite{itti2009bayesian}. We motivated our question format for eliciting surprise from Yang et al.'s previous work~\cite{yang2023subjective} and asked participants to drag a slider to report their degree of surprise. The slider ranged from 0 to 100, with three anchors on 0 (not at all surprised), 50 (to some extent surprised), and 100 (completely surprised). %

\paragraph{Trust}
We evaluated participants' perceived trust, defined as the confidence of each individual in the capabilities of the trustee (\ie visualization designs)~\cite{mcallister1995affect}. We used the same trust decomposition method (\textit{accuracy, fairness, reliability, trustworthiness, and understandability}) proposed by Yang et al.~\cite{yang2023swaying} to capture perceived trust. We captured participant responses using a %
bipolar, 7-point scale.%

\paragraph{Perceived Effort}
We captured participants' perceived effort completing our task with the given treatment using a digital version of the NASA-TLX test~\cite{castro2022examining, hart1988development, mckendrick2018deeper}. Participants clicked on each of six scales (\textit{mental, physical, temporal, failure, effort, frustration}) to describe their perceived effort on a 21-point scale. We analyzed the result by each scale following Galy et al.'s recommendation~\cite{galy2018measuring}.

\subsection{Procedure and Tasks}
\label{subsec:procedure-tasks}
The experiment consisted of three stages: tutorial scripts (Fig.~\ref{fig:experiment-procedure}, S1), main tasks (Fig.~\ref{fig:experiment-procedure}, S2), and closing questions (Fig.~\ref{fig:experiment-procedure}, S3). Before the experiment, participants completed an IRB-approved consent protocol. Then, we used Qualtrics.com to deliver the experimental survey and randomly assign participants to a visualization design group, forming the between-subject condition.

\bfsection{S1: Tutorial Scripts.}
At the start of the experiment, we briefed participants on the data context, tasks, and study procedure. We provided plot-specific instructions to introduce the visualization design they viewed (see supplementary material). %
After passing an attention check, participants were presented with a tutorial question to familiarize them with the interface of the two tasks that required clicking on the image (EQ1: Mortality Count Forecast, and EQ2: Mortality Range Forecast).

\bfsection{S2: Main Task Overview.}
During the main task, we randomly showed participants the repeated measures (\ie sub-datasets T2-6 shown in Tab.~\ref{tab:exp-dataset-choice}). Participants completed all plot evaluation tasks (S2.1 in Fig.~\ref{fig:experiment-procedure}) %
for each sub-dataset. %
Subsequently, participants reported their perceived effort (S2.2 in Fig.~\ref{fig:experiment-procedure}). Then, participants answered two open-ended questions (S2.3 in Fig.~\ref{fig:experiment-procedure}) about their strategies answering the range forecast question (EQ2 in Fig.~\ref{fig:experiment-procedure}), and the reasoning for their surprise or lack thereof when viewing actual outcomes (EQ5 in Fig.~\ref{fig:experiment-procedure}). We analyzed these open-ended responses to better understand the effects observed in S2.1 and S2.2 in Fig.~\ref{fig:experiment-procedure}.

\bfsection{S3: Closing Questions.}
We concluded the study with a short graph literacy survey~\cite{okan2019using} so we could control for individual differences in graph literacy as a covariate. We also collected participants' demographic data, including age, sex, and education level.

\subsection{Participants}
\label{subsec:participants}

We conducted both experiments online with participants from Prolific.com. Our target sample size was \textit{n} = 100 for each between-subject group in both studies based on a power analysis from a similar prior study~\cite{padilla2023multiple}. Our prescreening criteria dictated that participants should be over 18, fluent in English, reside in the US, and only enroll in one of the experiments. Evidenced by prior works~\cite{padilla2022impact,padilla2023multiple}, we argue the general public can meaningfully engage with COVID-19 forecast and uncertainty visualizations, exhibit systematic differences in judgments across visualization designs, and update their beliefs after viewing forecasts. Each experiment took about 20 minutes to complete, for which we paid at a rate of 16 USD per hour. 

We collected 711 responses for Experiment 1 (601, then 110 at a later time with the Horizon Sampled MFV condition). A screening process excluded 16 participants due to experimental setup issues and significant deviations in completion time (more than 2 standard deviations(SD) from the mean), leaving 695 qualified participants with an average age of 42.8 years (\statSD{15.4} years), among which 339 self-identified as female (48.8\%). Since Horizon Sampled MFVs were not tested concurrently with the other visualization designs, we observed a different mean age for this condition (\statMean{34.0} years, \statSD{11.7} years), signaling a potential population shift. We recruited 400 participants for Experiment 2. Identical postscreening criteria as Experiment 1 yielded 389 valid responses, among which 186 self-identified as female (47.8\%), and the average age was 35.5 years (\statSD{11.9} years).

\begin{table}[t]
\small
  \centering
  \caption{Key results from the omnibus analysis in Experiments 1 and 2. The first two rows in each experiment are the referent cases (\MFV{MFV}). We report the intercept $\beta$ (the degree of change between the compared conditions) and t or z value of each condition.}%
  \resizebox{.93\linewidth}{!}{
    \begin{tabular}{rrrrrr}
    \toprule
    \multicolumn{1}{l}{\multirow{2}[4]{*}{\textbf{Chart Type}}} & \multicolumn{1}{c}{\multirow{2}[4]{*}{\textbf{Stat}}} & \multicolumn{3}{c}{\textbf{Performance}} & \multicolumn{1}{c}{\multirow{2}[4]{*}{\textbf{Surprise}{~$\downarrow$}}} \\
\cmidrule{3-5}          &       & \multicolumn{1}{r}{Error{~$\downarrow$}} & \multicolumn{1}{r}{R.Acc.{~$\uparrow$}} & \multicolumn{1}{r}{Range} &  \\
    \midrule
    \multicolumn{6}{l}{\textbf{Experiment 1}} \\
    \midrule
    \multicolumn{1}{l}{\multirow{2}[1]{*}{\MFV{MFV}}} & \multicolumn{1}{c}{$\beta$} & \textcolor[HTML]{4B9C7A}{92.88} & \textcolor[HTML]{4B9C7A}{0.07} & \textcolor[HTML]{4B9C7A}{158.83} & \textcolor[HTML]{4B9C7A}{47.01} \\
          & \multicolumn{1}{c}{$t/z$} & \textcolor[HTML]{4B9C7A}{4.72} & \textcolor[HTML]{4B9C7A}{0.11} & \textcolor[HTML]{4B9C7A}{8.34} & \textcolor[HTML]{4B9C7A}{5.92} \\
    \multicolumn{1}{l}{\multirow{2}[0]{*}{\HMFV{Horizon MFV}}} & \multicolumn{1}{c}{$\beta$} & \cellcolor[rgb]{ .741,  .843,  .906}-14.73 & 0.32  & 5.01  & \cellcolor[rgb]{ .42,  .682,  .839}-9.02 \\
          & \multicolumn{1}{c}{$t/z$} & \cellcolor[rgb]{ .741,  .843,  .906}-2.94 & 1.50  & 0.42  & \cellcolor[rgb]{ .42,  .682,  .839}-3.69 \\
    \multicolumn{1}{l}{\PMFV{Prog. MFV}} & \multicolumn{1}{c}{$\beta$} & 7.40  & \cellcolor[rgb]{ .898,  .898,  .898}-0.52 & -4.44 & \cellcolor[rgb]{ .898,  .898,  .898}6.06 \\
    \multicolumn{1}{l}{\PMFV{(Frequency-mapped)}} & \multicolumn{1}{c}{$t/z$} & 1.48  & \cellcolor[rgb]{ .898,  .898,  .898}-2.45 & -0.38 & \cellcolor[rgb]{ .898,  .898,  .898}2.49 \\
    \multicolumn{1}{l}{\multirow{2}[0]{*}{\CI{CI}}} & \multicolumn{1}{c}{$\beta$} & 3.27  & \cellcolor[rgb]{ .741,  .843,  .906}-0.60 & -5.21 & \cellcolor[rgb]{ .898,  .898,  .898}5.73 \\
          & \multicolumn{1}{c}{$t/z$} & 0.66  & \cellcolor[rgb]{ .741,  .843,  .906}-2.85 & -0.44 & \cellcolor[rgb]{ .898,  .898,  .898}2.36 \\
    \multicolumn{1}{l}{\multirow{2}[0]{*}{\violin{Violin}}} & \multicolumn{1}{c}{$\beta$} & 3.77  & \cellcolor[rgb]{ .42,  .682,  .839}-0.93 & \cellcolor[rgb]{ .741,  .843,  .906}-33.35 & \cellcolor[rgb]{ .42,  .682,  .839}11.46 \\
          & \multicolumn{1}{c}{$t/z$} & 0.76  & \cellcolor[rgb]{ .42,  .682,  .839}-4.31 & \cellcolor[rgb]{ .741,  .843,  .906}-2.82 & \cellcolor[rgb]{ .42,  .682,  .839}4.71 \\
    \multicolumn{1}{l}{\multirow{2}[0]{*}{\density{Density}}} & \multicolumn{1}{c}{$\beta$} & \cellcolor[rgb]{ .898,  .898,  .898}10.37 & \cellcolor[rgb]{ .42,  .682,  .839}-0.98 & \cellcolor[rgb]{ .898,  .898,  .898}-24.7 & \cellcolor[rgb]{ .42,  .682,  .839}11.18 \\
          & \multicolumn{1}{c}{$t/z$} & \cellcolor[rgb]{ .898,  .898,  .898}2.09 & \cellcolor[rgb]{ .42,  .682,  .839}-4.58 & \cellcolor[rgb]{ .898,  .898,  .898}-2.10 & \cellcolor[rgb]{ .42,  .682,  .839}4.62 \\
    \multicolumn{1}{l}{\multirow{2}[1]{*}{\mean{Mean-only}}} & \multicolumn{1}{c}{$\beta$} & 4.23  & \cellcolor[rgb]{ .42,  .682,  .839}-1.74 & \cellcolor[rgb]{ .42,  .682,  .839}-68.44 & \cellcolor[rgb]{ .42,  .682,  .839}9.60 \\
          & \multicolumn{1}{c}{$t/z$} & 0.85  & \cellcolor[rgb]{ .42,  .682,  .839}-7.84 & \cellcolor[rgb]{ .42,  .682,  .839}-5.82 & \cellcolor[rgb]{ .42,  .682,  .839}3.97 \\
    \midrule
    \multicolumn{6}{l}{\textbf{Experiment 2}} \\
    \midrule
    \multicolumn{1}{l}{\multirow{2}[1]{*}{\MFV{MFV}}} & \multicolumn{1}{c}{$\beta$} & \textcolor[HTML]{4B9C7A}{91.49} & \textcolor[HTML]{4B9C7A}{0.02} & \textcolor[HTML]{4B9C7A}{159.69} & \textcolor[HTML]{4B9C7A}{46.5} \\
          & \multicolumn{1}{c}{$t/z$} & \textcolor[HTML]{4B9C7A}{4.84} & \textcolor[HTML]{4B9C7A}{0.05} & \textcolor[HTML]{4B9C7A}{6.01} & \textcolor[HTML]{4B9C7A}{5.93} \\
    \multicolumn{1}{l}{\multirow{2}[0]{*}{\HMFV{Horizon MFV}}} & \multicolumn{1}{c}{$\beta$} & \cellcolor[rgb]{ .898,  .898,  .898}-9.98 & 0.28  & 5.64  & \cellcolor[rgb]{ .42,  .682,  .839}-8.64 \\
          & \multicolumn{1}{c}{$t/z$} & \cellcolor[rgb]{ .898,  .898,  .898}-1.97 & 1.37  & 0.44  & \cellcolor[rgb]{ .42,  .682,  .839}-3.70 \\
    \multicolumn{1}{l}{\BMFV{Prog. MFV}} & \multicolumn{1}{c}{$\beta$} & -6.02 & \cellcolor[rgb]{ .42,  .682,  .839}0.81 & \cellcolor[rgb]{ .42,  .682,  .839}43.03 & \cellcolor[rgb]{ .741,  .843,  .906}-6.71 \\
    \multicolumn{1}{l}{\BMFV{(Base)}} & \multicolumn{1}{c}{$t/z$} & -1.18 & \cellcolor[rgb]{ .42,  .682,  .839}3.87 & \cellcolor[rgb]{ .42,  .682,  .839}3.36 & \cellcolor[rgb]{ .741,  .843,  .906}-2.86 \\
    \multicolumn{1}{l}{\multirow{2}[1]{*}{\violin{Violin}}} & \multicolumn{1}{c}{$\beta$} & 0.88  & \cellcolor[rgb]{ .898,  .898,  .898}-0.44 & 3.91  & \cellcolor[rgb]{ .42,  .682,  .839}8.02 \\
          & \multicolumn{1}{c}{$t/z$} & 0.17  & \cellcolor[rgb]{ .898,  .898,  .898}-2.14 & 0.31  & \cellcolor[rgb]{ .42,  .682,  .839}3.46 \\
    \midrule
          &    Legend: & \cellcolor[rgb]{ .42,  .682,  .839}$<0.001$ & \cellcolor[rgb]{ .741,  .843,  .906}$<0.005$ & \cellcolor[rgb]{ .898,  .898,  .898}$<0.05$ & $>0.05$ \\
    \bottomrule
    \end{tabular}%
    
}
  \vspace{-8px}
  \label{tab:results-full}%
\end{table}

\section{Results}
\label{sec:results}

To answer how different aggregation levels impact forecast readers' judgment, we quantitatively analyzed all 14 judgment-related measures (see Sec.~\ref{subsec:measures}). Below, we describe our quantitative analysis method (Sec.~\ref{result:method}) and report our findings (Sec.~\ref{result:exp1},~\ref{sec:res-exp2}). We further contextualize our findings by qualitatively analyzing the open-response questions through abductive coding (Sec .~\ref{subsec:results-open-response}).

\subsection{Quantitative Analysis Method} 
\label{result:method}

Since our study employed a mixed within- and between-subject design, we analyzed data using a linear-mixed effect method. We started with an omnibus analysis to estimate the variance accounted for by the experimental conditions (\textit{ChartType}) in each dependent variable (\textit{Measure}), in R notation~\cite{r2013r}:
\vspace{-4px}
\begin{equation}
        \text{\textit{Measure}} \sim \text{\textit{ChartType}} + \text{\textit{GraphLiteracy}} + (1 | \text{\textit{Time}}) + (1 | \text{\textit{ID}}) \nonumber
        \vspace{-4px}
\end{equation}
We used the \MFV{MFV} condition as the referent, comparing other visualization designs to it. We added \textit{graph literacy} as a covariate to account for readers' different visualization experiences. Additionally, we incorporated random intercepts for the forecast %
sub-datasets (repeated measures) and participant IDs to account for dataset differences and individual differences. We treated our 7-point Likert scale measures as continuous due to having more than five response categories~\cite{bauer2011fitting}. Moreover, because perceived effort was collected only once, we used a model without random intercepts.%

In \texttt{R}, we fitted mixed-effects models using the \texttt{lmerTest} package v3.1.3 and conducted pair-wise comparisons with the \texttt{emmeans} package v1.10.0. %
We report $\beta$, \textit{t}, \textit{p} values; the confidence interval range \textit{CI}; and the degree of freedom for the effects of each metric. Data and quantitative analysis script are in an OSF repository\footnote{\url{https://osf.io/vc3kn/}}. Key results are summarized in Table~\ref{tab:results-full}.

\subsection{Experiment 1 Results}
\label{result:exp1}

\begin{figure*}[t]
    \centering
    \begin{minipage}[t]{0.65\linewidth}
        \centering
        \includegraphics[width=1\linewidth]{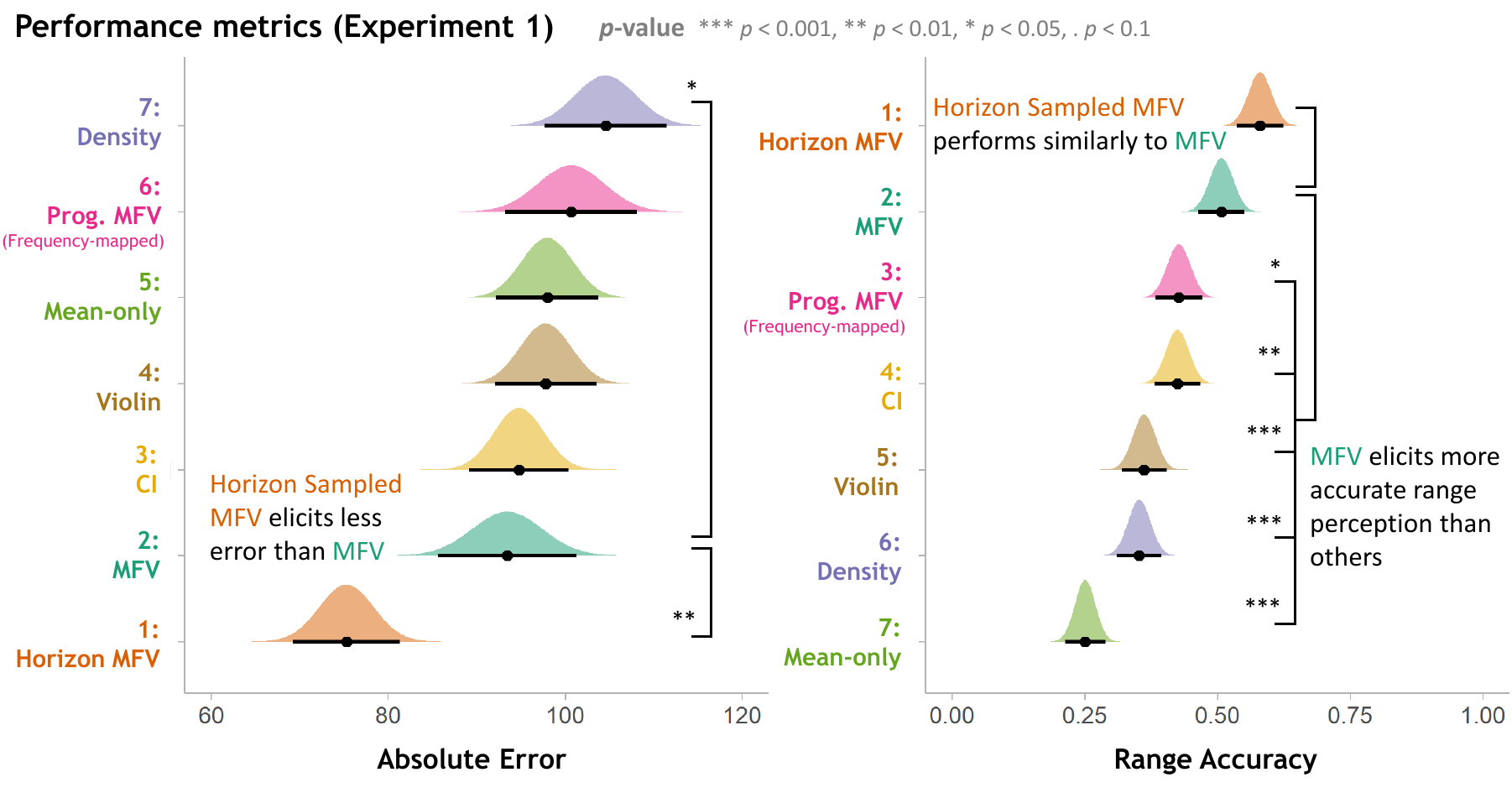}
        \caption{Illustration of the response distribution for two performance metrics in Experiment 1, with annotations of key findings.}
          \vspace{-8px}
        \label{fig:result-performance}
    \end{minipage}
    \hfill
    \begin{minipage}[t]{0.325\linewidth}
        \centering
        \includegraphics[width=1\linewidth]{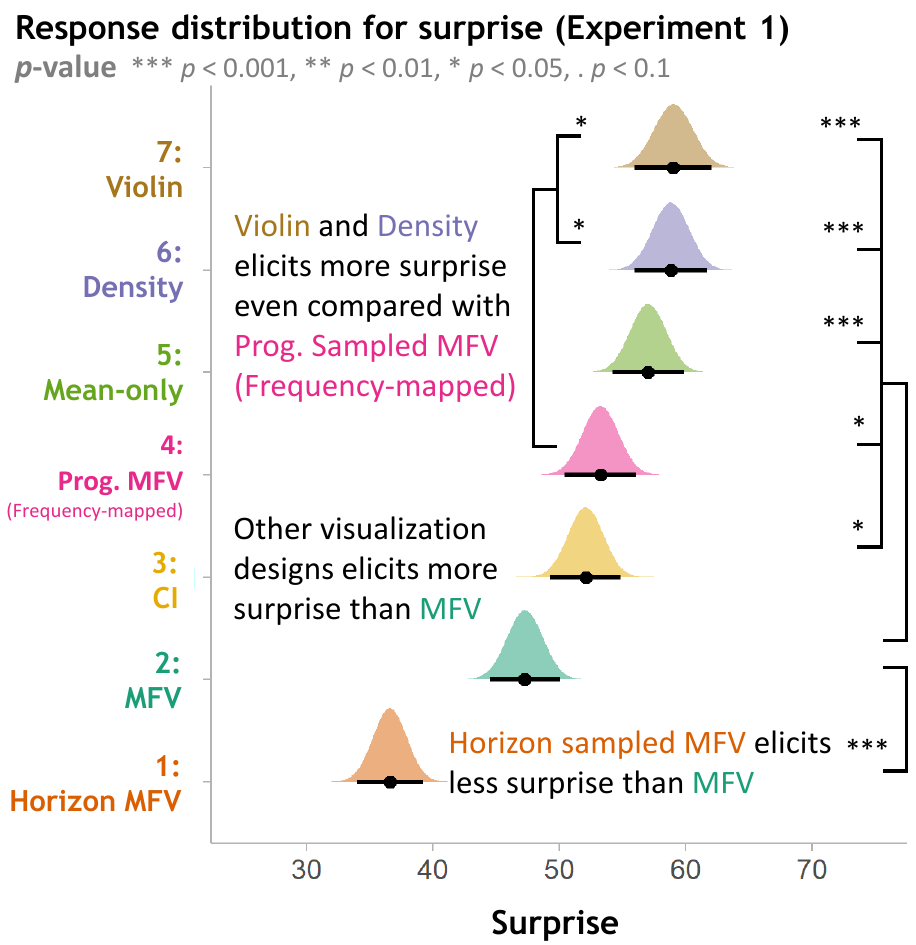}
        \caption{Illustration of the response distribution of reported surprise in Experiment 1.}
          \vspace{-8px}
    \label{fig:result-surprise-exp1}
    \end{minipage}
\end{figure*}

\textbf{Performance.}
\textbf{Participants performed best with the \HMFV{Horizon Sampled MFV}} (see Fig.~\ref{fig:result-performance} and Tab.~\ref{tab:results-full} Experiment 1, Performance columns). Comparing \HMFV{Horizon Sampled MFV} to the referent \MFV{MFV}, participants showed significantly less absolute error (\statBeta{-14.73}, \statT{3464}{-2.94}, \statP{0.003}, \statCI{-24.56}{-4.91}) while maintaining a similar range accuracy (\statBeta{0.32}, \statZ{1.50}, \statP{0.133}, \statCI{-0.10}{0.74}) and range estimate (\statBeta{5.01}, \statT{3464}{0.42}, \statP{0.674}, \statCI{-18.31}{28.33}). Such results indicate that \HMFV{Horizon Sampled MFV} elicited better judgments from participants, which are not a result of perceiving a wider range of forecasts.

Meanwhile, participants performed significantly worse with \MFV{MFVs} than with \density{Density} Plots in both absolute error (\statBeta{10.37}, \statT{3464}{2.09}, \statP{0.037}, \statCI{0.65}{20.09}) and range accuracy (\statBeta{-0.98}, \statZ{-4.58}, \statP[\textless]{0.001}, \statCI{-1.40}{-0.56}). Participants performed similarly in absolute error in other visualization designs (\PMFV{Frequency-mapped Progressively Sampled MFV}, \mean{Mean-only}, \violin{Violin}, and \CI{CI}) compared to \MFV{MFV}, whereas their range accuracy was significantly worse. Such results indicate a worse judgment on the forecast range using the above visualization designs.

Following the initial analysis using the omnibus model with the \MFV{MFV} as the referent condition, we conducted exploratory pairwise comparisons with Tukey adjustments to reveal differences along the aggregation level spectrum. Participants viewing the \HMFV{Horizon Sampled MFV} exhibited both significantly lower absolute error and significantly higher range accuracy than all other visualization designs, except for the \MFV{MFV}. Taking the auxiliary range estimate measure into perspective, we found that participants' tendency to make more accurate range estimates with \HMFV{Horizon Sampled MFV} is reliable. We found no significant pair-wise difference in range estimates between \HMFV{Horizon Sampled MFV} (\statMean{163.8}) with \MFV{MFV}, \CI{CI}, and \density{Density}, although \HMFV{Horizon Sampled MFV's} range estimate was significant compared to \violin{Violin} (\statMean{125.5}, \statP{0.024}) and \mean{Mean-only} (\statMean{90.4}, \statP[\textless]{0.001}) plots.

\smallskip
\noindent \textbf{Surprise. }
\textbf{Participants were least surprised viewing the \textbf{\HMFV{Horizon Sampled MFV}}} (see Fig.~\ref{fig:result-surprise-exp1} and Table~\ref{tab:results-full}, Surprise). Participants were significantly less surprised viewing the \HMFV{Horizon Sampled MFV} (\statBeta{-9.02}, \statT{3464}{-3.69}, \statP[\textless]{0.001}, \statCI{-13.82}{-4.23}) than \MFV{MFV} (\statMean{47.01}, \statCI{31.45}{62.57}). Participants felt significantly more surprised viewing all other visualization designs than \MFV{MFV}. A subsequent pair-wise analysis with Tukey adjustments further revealed that participants viewing \HMFV{Horizon Sampled MFVs} felt significantly less surprised than those viewing all other visualization designs, except for \MFV{MFVs}.

\smallskip
\noindent \textbf{Trust \& Effort.}
\textbf{Participants reported similar levels of trust and effort viewing most visualization designs compared to \MFV{MFVs}.} Within trust, participants only reported a significant difference between \violin{Violin} Plots and \MFV{MFVs} in fairness (\statBeta{0.31}, \statT{3464}{2.00}, \statP{0.045}, \statCI{0.01}{0.62}) and understandability (\statBeta{0.59}, \statT{3464}{3.03}, \statP{0.002}, \statCI{0.21}{0.97}). They also reported that \mean{Mean-only} plots were more understandable (\statBeta{0.47}, \statT{3464}{2.43}, \statP{0.015}, \statCI{0.09}{0.85}) than \MFV{MFVs}. The two designs eliciting the extremes of trust, based on the average trust ranking across five submeasures, were \violin{Violin} Plots (most trusted) and \MFV{MFVs} (least trusted).

For effort, \textbf{participants reported no significant difference between other designs and \MFV{MFVs}}, except \HMFV{Horizon Sampled MFVs}. Participants reading \HMFV{Horizon Sampled MFVs} reported a significant reduction in perceived effort (\statBeta{-2.81}, \statT{687}{-3.77}, \statP[\textless]{0.001}, \statCI{-4.28}{-1.34}), mental demand (\statBeta{-1.65}, \statT{687}{-2.26}, \statP{0.024}, \statCI{-3.09}{-0.22}), and frustration (\statBeta{-2.65}, \statT{687}{-3.65}, \statP[\textless]{0.001}, \statCI{-4.08}{-1.23}) than those reading \MFV{MFVs}.

However, because \HMFV{Horizon Sampled MFVs} was tested at a different time than other designs, we cannot rule out the possibility that a shifting participant population has introduced a confounding variable. Hence, we examine the reliability of all findings related to \HMFV{Horizon Sampled MFVs} in Experiment 2. %

\subsection{Experiment 2}
\label{sec:res-exp2}

As described in Sec.~\ref{subsec:experiment-design}, Experiment 2's main purpose was to determine the reliability of Experiment 1's \HMFV{Horizon Sampled MFV} results, and to further evaluate the effect of trajectory aggregation using \BMFV{Base Progressively Sampled MFVs}. In the following sections, we compare the results of Experiment 1 and 2, and provide an analysis of \BMFV{Base Progressively Sampled MFVs'} performance.

\subsubsection{Experiment 1 Results Reliability Check}
\label{subsec:results-reliability}

\begin{figure*}[t]
    \centering
    \begin{minipage}[t]{0.65\linewidth}
        \centering
        \includegraphics[width=1\linewidth]{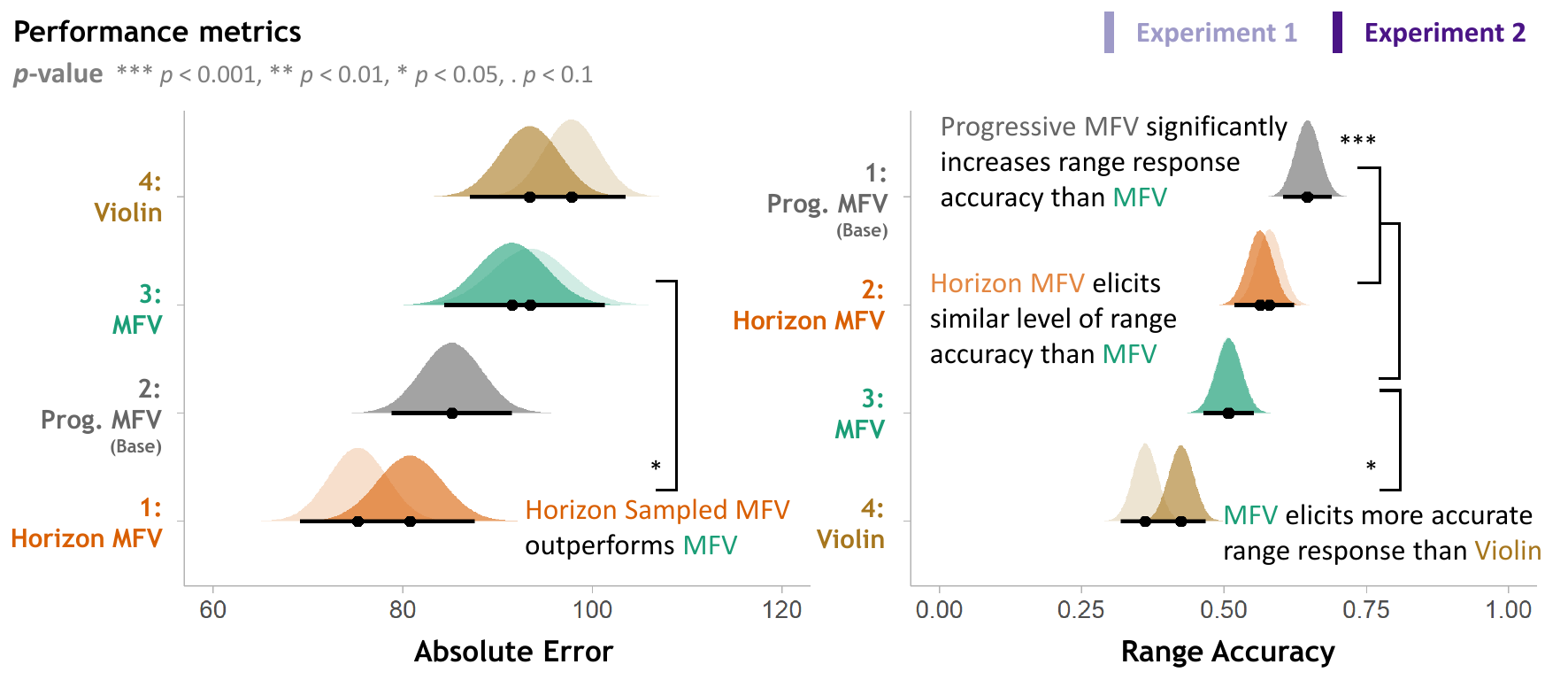}
        \caption{Response distribution for performance metrics comparing Experiment 1 and 2. Experiment 1 results are shown with lighter versions of the colors used in Experiment 2.}%
          \vspace{-8px}
        \label{fig:result-performance-exp2}
    \end{minipage}
    \hfill
    \begin{minipage}[t]{0.325\linewidth}
        \centering
        \includegraphics[width=1\linewidth]{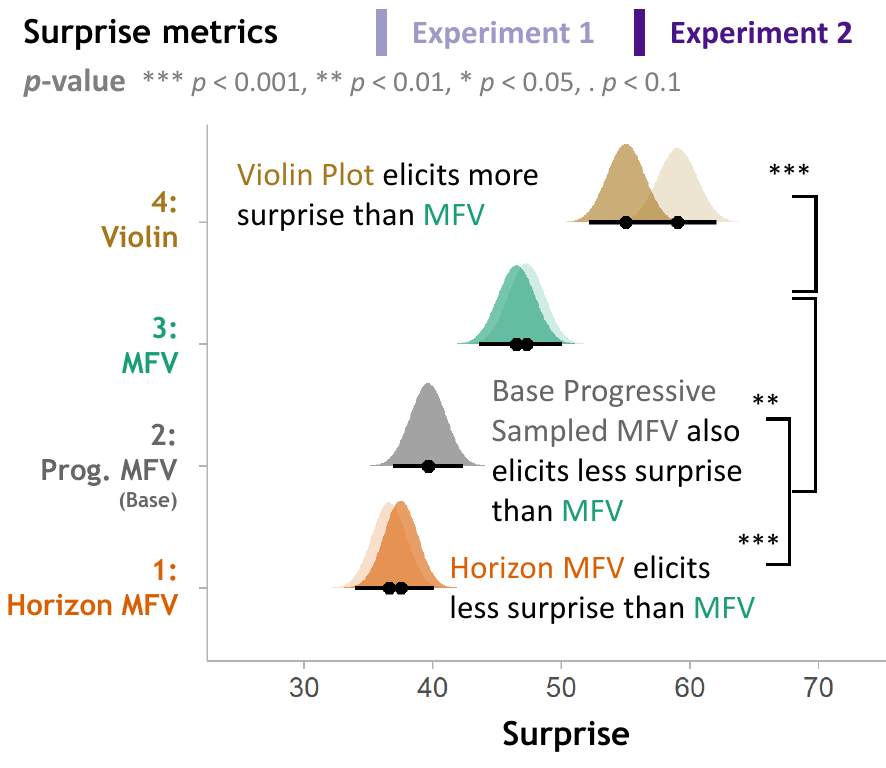}
        \caption{Comparison of surprise in Experiments 1 and 2.}%
          \vspace{-8px}
        \label{fig:result-surprise-exp2}
    \end{minipage}
\end{figure*}

\textbf{Performance.}
Through the same analysis process as Experiment 1, Experiment 2 provides additional support for previously found judgment performance results (Fig.~\ref{fig:result-performance-exp2} and Tab.~\ref{tab:results-full}, Experiment 2, Performance). These results suggest that participants' forecast estimates were significantly better when reading \HMFV{Horizon Sampled MFVs} (\statBeta{-9.98}, \statT{1937}{-1.97}, \statP{0.049}, \statCI{-19.91}{-0.05}) than \MFV{MFVs}. Meanwhile, participants who read these designs performed similarly in estimating forecasts' range (\statBeta{0.28}, \statZ{1.37}, \statP{0.171}, \statCI{-0.12}{0.68}). 

The \textit{p}-values for error were close to our significance threshold \textit{p} = 0.05. Thus, we conducted a follow-up analysis with stricter exclusion criteria to verify the reliability of our results. Our experiment included an attention check in which participants completed a sentence with the word ``charts.'' Although some incorrectly wrote other answers (\eg ``cases''), they were initially included with our lenient exclusion criterion (starting with letter `c'). 
Applying a stricter criterion (accepting only ``chart'' or ``charts''), we arrive at a lower, more reliable \textit{p}-value of 0.032. %
Still, below we report the more conservative results that include all participants who passed our more lenient attention check criteria. We advise readers to interpret our findings in this context.

Visual analysis of the response distribution for all judgment performance metrics (Fig.~\ref{fig:result-performance-exp2}) also reveals cross-experiment consistency for \MFV{MFVs} and \HMFV{Horizon Sampled MFVs}. However, the performance results for \violin{Violin} Plots show a higher variance than the other conditions we retested. Although we replicate participants' response that \violin{Violin} Plots are comparable (\statBeta{0.88}, \statT{1937}{0.17}, \statP{0.862}, \statCI{-9.01}{10.76}) to \MFV{MFVs} in error and significantly worse (\statBeta{-0.44}, \statZ{-2.14}, \statP{0.033}, \statCI{-0.84}{-0.04}) than \MFV{MFVs} in range accuracy, we observes that, in Experiment 2, participants who read \violin{Violin} Plots (\statMean{164}) selected a wider range than those in Experiment 1 (\statMean{125.5}). Despite this broader range, Experiment 2's \violin{Violin} Plots' range accuracy remains consistent with Experiment 1's, indicating that participants made the least accurate judgments using \violin{Violin} Plots. We also confirm through repeated experiments that \HMFV{Horizon Sampled MFVs} reliably lead to better judgment performance than \MFV{MFVs} and \violin{Violin Plots}.

\smallskip
\noindent \textbf{Surprise.}
In addition to performance, we also validated Experiment 1's surprise metric results (Tab.~\ref{tab:results-full}, Experiment 2, Surprise). Participants were again significantly less surprised viewing \HMFV{Horizon Sampled MFVs} (\statBeta{-8.64}, \statT{1937}{-3.70}, \statP[\textless]{0.001}, \statCI{-13.21}{-4.06}) in comparison to \MFV{MFVs}, and participants were significantly more surprised viewing \violin{Violin} Plots (\statBeta{8.02}, \statT{1937}{3.46}, \statP[\textless]{0.001}, \statCI{3.47}{12.58}) in comparison to \MFV{MFVs}. Visual analysis of both experiments' response distribution (Fig.~\ref{fig:result-surprise-exp2}) displays cross-experiment consistency.

\smallskip
\noindent \textbf{Trust \& Effort.}
\textbf{Overall, participants reported similar levels of trust and effort viewing all visualization designs.} Regarding trust, while we were able to replicate participants' responses that \HMFV{Horizon Sampled MFV} and \violin{Violin} incurred similar trust to \MFV{MFV}, we did not repeat the other significance regarding the \violin{Violin} Plot. Regarding effort, participants reported similar levels of effort when viewing all the visualization designs tested in Experiment 2. We speculate that the significant results for the \HMFV{Horizon Sampled MFV} in perceived effort, mental demand, and frustration were likely the effect of the population shift.

In summary, Experiment 2 confirmed key findings of Experiment 1. We confirmed that the improvement in judgment performance and the decrease in surprise for \HMFV{Horizon Sampled MFV} are reliable, and the decreased effort of \HMFV{Horizon Sampled MFV} was likely caused by the population shift. Meanwhile, we discovered a recurring theme indicating that the response for \violin{Violin} Plots was more volatile than \MFV{MFV} or the \HMFV{Horizon Sampled MFV}. %

\subsubsection{Base Progressively Sampled MFV Analysis}
\label{subsec:results-hmfv}
We included \BMFV{Base Progressively Sampled MFV} in Experiment 2 to explore if the excessive visual encoding in \PMFV{Progressively Sampled MFV (Frequency-mapped)} impacted the evaluation result. Participants did not rate the \PMFV{Progressively Sampled MFV (Frequency-mapped)} as good as the \HMFV{Horizon Sampled MFV} in Experiment 1 despite being more information-rich. Qualitative coding open-ended responses hinted that participants likely fixated on the darker lines (Sec.~\ref{subsec:results-open-response}). Thus, we removed gray-scale mapping and created the \BMFV{Base Progressively Sampled MFV} to further evaluate the reliability of the results in Experiment 1.

Results revealed that participants viewing \BMFV{Base Progressively Sampled MFV} made significantly more accurate range estimate (\statBeta{0.81}, \statZ{3.87}, \statP[\textless]{0.001}, \statCI{0.40}{1.22}) than \MFV{MFV}. However, the improved range estimate performance came at  the cost of using a considerably larger range estimate (\statBeta{43.03}, \statT{1937}{3.36}, \statP[\textless]{0.001}, \statCI{17.95}{68.11}). Participants were also significantly less surprised viewing the \BMFV{Base Progressively Sampled MFV} (\statBeta{-6.71}, \statT{1937}{-2.86}, \statP{0.004}, \statCI{-11.30}{-2.11}) than \MFV{MFV}. However, the trade-off between improved judgment performance and surprise was a decrease in trust. Participants consistently rated \BMFV{Base Progressively Sampled MFV} last in trust among the visualization designs tested in Experiment 2. Specifically, participants rated trustworthiness (\statBeta{-0.33}, \statT{1937}{-2.22}, \statP{0.027}, \statCI{-0.63}{-0.04}) and accuracy (\statBeta{-0.32}, \statT{1937}{-2.19}, \statP{0.028}, \statCI{-0.61}{-0.03}) trust submeasures significantly worse than \MFV{MFV}. Effort metrics were similar to \MFV{MFV}.

\infobox{Summary of findings}{Across Experiments 1 and 2, \HMFV{Horizon Sampled MFV} (\ie aggregating on the selected level) consistently supported participants in accurately completing prediction tasks and minimizing surprise when presented with actual outcomes. Furthermore, participants using the \HMFV{Horizon Sampled MFV} were more likely to anticipate a range of outcomes that accurately encompassed the actual result.}

\subsection{Open Response Questions}
\label{subsec:results-open-response}

\begin{table}[t]
\small
  \centering
  \caption{Percentage of participants who reported using visual forecast cue strategy to extrapolate outcome range.}
    \begin{tabular}{lrrr}
    \toprule
    \multirow{2}[2]{*}{\textbf{Visualization Design}} & \multirow{2}[2]{*}{\textbf{\% Strategy}} & \multicolumn{2}{c}{\textbf{Inter-rater Statistics}} \\
          &       & \multicolumn{1}{l}{\boldmath{}\textbf{Cohen's $\kappa$}\unboldmath{}} & \multicolumn{1}{l}{\textbf{\% Agreement}} \\
    \midrule
    \MFV{MFV}   & 0.58  & 0.90  & 0.95  \\
    \PMFV{Prog. MFV (Frequency)} & 0.58  & 0.85  & 0.93  \\
    \HMFV{Horizon MFV} & 0.55  & 0.94  & 0.97  \\
    \BMFV{Prog. MFV (Base)} & 0.53  & 0.89  & 0.95  \\
    \CI{CI}    & 0.48  & 0.62  & 0.81  \\
    \violin{Violin} & 0.39  & 0.96  & 0.98  \\
    \density{Density} & 0.37  & 0.96  & 0.98  \\
    \mean{Mean-only}  & 0.25  & 0.84  & 0.95  \\
    \bottomrule
    \end{tabular}%
  \label{tab:result-qualitative}%
  \vspace{-12px}
\end{table}%

To better understand the quantitative findings, we qualitatively examined the open-response questions through abductive coding. The goals of our analysis were to find possible explanations for 1) why some designs elicited better judgment utility and 2) why participants displayed a more consistent response pattern in some visualization designs (\eg \MFV{MFV}) than others (\eg \violin{Violin} Plot).

To analyze the open responses, we followed an abductive coding procedure. Two coders first inductively coded the range strategy question, which contained indications of participants' reading strategy. We observed that participants' inclination to use ``forecast cues'' (\ie use information in the purple box in Fig.~\ref{fig:stimuli-sample}, coded as \textit{visual forecast cue}) might be contributing factors for both performance and consistency. The two coders then discussed what qualified as a \textit{visual forecast cue}. Specifically, we determined any response that explicitly or implicitly mentioned ``forecasts'' or ``predictions'' as those who used the \textit{visual forecast cues}. To ensure clarity, we excluded the vague responses (\eg \textit{``I tried to follow the trend''}) or those that explicitly used other strategies (\eg \textit{``I looked at the history of the chart''} or \textit{``I did my best educated guess''}). Both coders subsequently coded all 790 responses from eight visualization designs (all responses from Experiment 1 except \BMFV{Base Progressive Sampled MFV} from Experiment 2).

Table~\ref{tab:result-qualitative} shows the percentage of participants for each visualization design that used the visual forecast cue strategy. To verify both coders were consistent in coding, we computed Cohen's kappa and percent agreement using \texttt{irr} package v0.84.1 for inter-rater reliability, and reported both values as advised by McHugh~\cite{mchugh2012interrater}. 

Our qualitative analysis indicated that participants were more inclined to use the visual forecast cue 
when provided with \MFV{MFV} or its sampled variants than other visualization designs. Given the information anchor provided by the visualizations, it is likely that response consistency is related to the inclination to use forecast cues when judging future outcomes. We hypothesize that there could also be a connection between the visual forecast cue strategy and judgment performance, as the \HMFV{Horizon Sampled MFV}, \BMFV{Base Progressively Sampled MFV}, and \MFV{MFV} have both high judgment performance and usage of the visual forecast cue strategy. The exception that contrasts this hypothesis is the \PMFV{Frequency-mapped Progressively MFV}, which indicates that the more complex frequency mapping design impacted participants' judgments. The design for frequency mapping  (\ie using depth of line segments to present frequency) possibly overtly drew participants' attention to those segments: \textit{``Knowing that the darker lines had more consensus among the models than lighter lines, I used the upper and lower darkest lines to establish my range. I used those points assuming the more consensus would mean the higher probability the actual result would land somewhere between.''} The above finding also motivated the \BMFV{Base Progressively Sampled MFV} design as we sought to reduce the bias associated with the darker lines, which we observe improvements in judgment performance.

\section{Discussion}
\label{sec:discussion}

Through Experiments 1 and 2, we examined different aggregation levels on multiple forecasts (represented by eight visualization designs), impact judgment under uncertainty. A key finding is that the Horizon Sampled MFV (\ie aggregating on the level of selection) is likely the most effective in enabling readers to accurately judge future events, thereby reducing their surprise upon encountering real outcomes. Additionally, participants viewing the Horizon Sampled MFV were more likely to anticipate a range of outcomes that accurately included the actual outcome. 

However, %
readers did not identify one single approach that outperformed all others. We observe that designs with minimal aggregations are likely more consistent perceived across populations. Additionally, designs that lead readers to focus on forecasts themselves generally exhibited greater consistency. %

Our findings extend the understanding of visualizing multiple forecasts by grounding the design choices in the trade-off between expressiveness and effectiveness. We extend findings comparing two extremes of aggregation (\ie no aggregation \vs full aggregation)~\cite{sarma2025more} by showing that partial aggregation (instantiated by the Horizon Sampled MFV) could potentially bring benefits in eliciting accurate judgments. We thus open a wider design space for designers and researchers to explore.

More broadly, we add to a growing body of evidence showing that distributional representations of uncertainty outperform summary-based or uncertainty-omitting visualizations~(\eg \cite{joslyn2013error, correll2014error, castro2022examining, kale2021visual}). These findings are particularly pertinent to crisis communication cases such as COVID-19 forecasting, where approximately 60\% of visualizations still rely on confidence intervals, despite evidence favoring designs using distributional representations~\cite{zhang2021mapping}. %

\subsection{Design Recommendations}

Synthesizing from our study findings, we recommend that visualization designers choose designs based on specific communication goals and prior knowledge of the forecasts:

\smallskip
\noindent
\textcolor{darkgrey}{\small\ding{108}} \textbf{Robust Forecast Communication}: If no specific communication goal is provided, we suggest aggregating by selecting a subset of forecasts (\eg using the Horizon Sampled MFV) because it is a highly comprehensive design and is consistent across populations and time points.  Participants viewing the Horizon Sampled MFV were also less surprised when actual outcomes were revealed--a feature that may help build long-term trust.

\smallskip
\noindent
\textcolor{darkgrey}{\small\ding{108}} \textbf{Stable Forecasting Communication}: %
All designs performed well in stable prediction scenarios (T4), where most forecasts closely aligned with the actual outcome. This is expected, as viewers can accurately predict the outcome by simply relying on the mean when it aligns perfectly with the actual result. Hence, if forecasts are known to be accurate in reflecting future outcomes, we recommend that designers choose an aggregation level based on other design considerations (\eg cognitive load, specific reader needs).

\smallskip
\noindent
\textcolor{darkgrey}{\small\ding{108}}  \textbf{Elicit Trust}: If ensuring trust in communication is the top priority, we recommend using more aggregated versions (\eg Violin Plot or Confidence Interval Plot). Although our results revealed mixed findings for trust ratings, the Violin Plot or Confidence Interval Plot generally recorded a high trust response. This finding aligns with prior results indicating that confidence intervals and displays without uncertainty are more trusted than MFVs~\cite{padilla2023multiple}. Additionally, the simplicity~\cite{elhamdadi2024vistrust} and the popularity of Violin and Confidence Interval Plots likely enhance their perceived trustworthiness.

\subsection{Limitations and Future Work}

There are several limitations of our work that inform future research opportunities. First, our study only evaluated the visualizations on a single dataset with a single context (disease forecast) with five sub-datasets. Although we intentionally extracted various truth-forecast relationships to balance ecological validity and experimental control, our results may be limited to other datasets with similar attributes to our five test cases. We also adopted a summative approach to the three dataset types (aligned, contrast, and outlier), leaving the specific effects of each type as an open question for future investigation. Future research should expand our study to broader datasets and scenarios and breakdown into dataset types to further evaluate the validity and expand of our findings.

Second, the measures we picked might not be comprehensive enough to capture the complex process of judgment. For example, other trust measurements (\eg~\cite{elhamdadi2024vistrust}) exist that might
reveal further insight regarding trust in the visualization designs. Our evaluation of surprise is also general, which might not be enough to uncover what contributes to the different surprise levels. Future research should seek to focus on one of the measures towards understanding how humans perceive those visualizations.

Another limitation of our work stems from the crowd-sourced data collection process. Although we checked the consistency between Experiments 1 and 2 and reported the agreement and disparity, the population on Prolific might have distinct attributes that limit its generalizability. While we found consistency across Experiments 1 and 2, there are several factors that might impact the generalizability of our findings. For example, the consistency was found in populations from the US who are fluent in English. Participants from other countries, speaking languages other than English, might have different cognitive processes than those in the US. Future work should seek to understand whether, if so, how different demographics interpret visualizations of multiple forecasts.

While MFVs are effective at conveying distributional information, prior work suggests an upper bound of approximately six to nine trajectories, beyond which overplotting increases and trade-offs between trust and performance emerge~\cite {padilla2023multiple}.  Our findings align with this constraint and demonstrate that partial aggregation can serve as an effective strategy for preserving key distributional information without incurring the overplotting associated with large numbers of forecasts. However, our study evaluated a base set of forecasts from 24-44 models. For datasets with substantially larger forecasts or higher-dimensional uncertainty, additional aggregation or interaction techniques would likely be required.

Lastly, our design space for visualizing multiple forecasts was not exhaustive. While we focused on the aggregation level, we did not evaluate how visual encodings such as color, line thickness, or opacity interact with aggregation to influence judgment. Encoding aggregation amount--for example, by using thicker or darker lines to indicate forecasts supported by many models--may help readers interpret aggregation levels more accurately, but could also introduce attentional biases that lead viewers to overweight emphasized trajectories and discount the broader range of possible outcomes~\cite{kale2021visual}. 
Future work should systematically explore these trade-offs to better understand how visual encoding and aggregation jointly shape uncertainty communication.

\section{Conclusion}
\label{sec:conclusion}

The prevalence of multiple forecasts has ignited the need for effective visualization techniques to communicate their results. This study advances prior work on multiple forecast visualizations for time-series data by evaluating how different levels of aggregation across multiple forecasts affect judgment performance. Through two large-scale online comparative experiments, we evaluated how various uncertainty visualization techniques for multiple forecasts impacted judgment performance across five time points and 14 measures. Our findings highlight the value of striking a balance. We found that the selected aggregation level, in the form of Horizon Sampled MFV, is particularly effective in improving judgment performance and reducing surprise when the actual outcome is revealed. However, in practice, choosing the optimal visual design requires careful consideration of communication goals since there is no one-size-fits-all approach that excels across all metrics.


\acknowledgments{
This work was supported in part by NSF Grant \#2428149, NIH Grant \#1R01AI188576-01, and a Northeastern University Tier-1 Research Seed Grant. We want to thank all participants for taking part in our study. Additionally, we extend our thanks to the anonymous reviewers for their constructive feedback.
}

\bibliographystyle{abbrv-doi}

\balance
\bibliography{
bib/new_additions, bib/website_ref, bib/zotero_export
}

\end{document}